\documentclass[%
preprint,
 amsmath,amssymb,
 aps,
]{revtex4-2}

\usepackage{graphicx}
\usepackage{dcolumn}
\usepackage{bm}
\usepackage{xcolor}

\begin{document}


\title{Generalization of the Gauss Map: \\ 
A jump into chaos with universal features}

\author{Christian Beck}
\email{c.beck@qmul.ac.uk}
\affiliation{
Centre for Complex Systems, School of Mathematical Sciences, Queen Mary University of London, London E1 4NS, UK 
}%

\author{Ugur Tirnakli}
\email{ugur.tirnakli@ieu.edu.tr}
\affiliation{
Department of Physics, Faculty of Arts and Sciences, Izmir University of Economics, 35330, Izmir, Turkey
}%

\author{Constantino Tsallis}
\email{tsallis@cbpf.br}
\affiliation{%
Centro Brasileiro de Pesquisas Fisicas and National Institute of Science and Technology for Complex Systems, Rua Dr. Xavier Sigaud 150, 22290-180, Rio de Janeiro, Brazil\\
Santa Fe Institute, 1399 Hyde Park Road, 87501, Santa Fe, USA\\
Complexity Science Hub Vienna, Josefstadter Strasse 39, 1080, Vienna,
Austria
}%

\begin{abstract}
The Gauss map (or continued fraction map) is an important dissipative
one-dimensional discrete-time dynamical system that exhibits chaotic behaviour and which generates a symbolic dynamics consisting of infinitely many different symbols. Here we introduce a generalization of the Gauss map which is given by
$x_{t+1}=\frac{1}{x_t^\alpha} - \Bigl[\frac{1}{x_t^\alpha} \Bigr]$
where $\alpha \geq 0$ is a parameter and $x_t \in [0,1]$ ($t=0,1,2,3,\ldots$).
The symbol $[\dots ]$ denotes the integer part. This map reduces to the
ordinary Gauss map for $\alpha=1$. 
The system exhibits a sudden `jump into chaos' at the critical parameter value 
$\alpha=\alpha_c \equiv 0.241485141808811\dots$ which we analyse in detail in this paper. 
Several analytical and numerical results are established for this new map as a function of the 
parameter $\alpha$. In particular, we show that, at the critical point, the invariant density 
approaches a $q$-Gaussian with $q=2$ (i.e., the Cauchy distribution), which becomes infinitely 
narrow as $\alpha \to \alpha_c^+$. Moreover, in the chaotic region for large values of the 
parameter $\alpha$ we analytically derive approximate formulas for the invariant density, 
by solving the corresponding Perron-Frobenius equation. For $\alpha \to \infty$ the uniform 
density is approached. 
We provide arguments that some features of this transition scenario are universal and are 
relevant for other, more general systems as well. 
\end{abstract}

\maketitle


\newpage

\section{Introduction}
The famous Gauss map (or continued fraction map) given by $x_{t+1}=f(x_t) = 1/x_t$ (mod 1) on the unit interval $[0,1]$ is a mathematically very interesting discrete-time dynamical system with many applications 
\cite{adler, sinai, corless, banchoff, BS}. It plays an important role in ergodic theory and number theory since it generates a symbolic dynamics given by the continued fraction expansion of the initial value $x$. It has positive Lyapunov exponent and exhibits
chaotic and mixing behaviour. The invariant density is explicitly known, already since the days of Gauss. 
The approach to equilibrium states (the second-largest eigenvalue of the Perron-Frobenious operator) has been studied, e.g. in \cite{mayer-roepstorff}. There are several known physical applications for this map, for example in general relativity for certain types of cosmologies that exhibit chaotic behavior such as the mixmaster cosmology \cite{barrow, chernoff} or for chaotic renormalization flows \cite{lacasa}. Generally, the Gauss map is a function from an oriented surface in Euclidean space to the unit sphere, which associates to every point on the surface its oriented unit normal vector. This emphasizes the importance of this map in general relativity and differential geometry.

In this paper we generalize and extend the original concept of the Gauss map. We will study a new one-parameter generalization of the Gauss map which we call
$\alpha$-Gauss map. It is given by $f(x) =1/x^\alpha$ (mod 1) on the interval $[0,1]$ for $\alpha \geq 0$. It reduces to the ordinary Gauss map for $\alpha =1$. Our main subject of interest is the
behaviour of this new map as a function of the parameter $\alpha$.

We find an interesting transition scenario  for this new map.
It  can be briefly summarized as a single sudden `jump into chaos' at a critical parameter value $\alpha_c\equiv 0.241485141808811$.... The phenomenon has a certain similarity with an interior crisis \cite{grebogi, ott},
sometimes also called an `explosive bifurcation' \cite{thomson, yue,luo}. 
However, in our case it occurs in an isolated way and it is the only
topological attractor change observed if $\alpha$ is varied.
The $\alpha$-Gauss map does not exhibit any period-doubling or tangent bifurcations or quasi-periodic routes or more complicated
crisis scenarios---there is just a single transition from a stable period-1 orbit
directly into the chaotic attractor, and chaos is then persistently observed for any 
$\alpha > \alpha_c$, including very large values of $\alpha$. 
At the critical point $\alpha_c$ we observe that the invariant density approaches an infinitely narrow $q$-Gaussian with $q=2$ which  eventually degenerates into a Dirac-delta function. Several aspects of the transition can be fully understood through  analytic means, and will be described in detail in the following sections.

While our 1-parameter generalization of the Gauss map is new and has not been 
studied before, related phenomena such as the sudden jump into chaos or the persistence of 
chaos above a critical threshold (without interruption by periodic windows) have been previously 
reported in the literature, though for very different classes of maps \cite{1,2,3,4,5,6,7,8,9,10,11}. 
Direct transitions from a stable period-1 orbit to chaos have been numerically observed 
in \cite{1,2} and a direct transition from a stable period-2 orbit into chaos in \cite{3}. 
The persistence of chaotic attractors without any interrupting periodic windows has been 
previously investigated in  \cite{7,8,9,10} (though for different types of maps) as well as 
for continuous time dynamical systems in \cite{11}. This phenomenon is often referred to as 
{\em robust chaos} \cite{8,9}.


Different from these previous papers, our main emphasis in the current paper is 
on the {\em invariant densities} that are observed at the transition point, as well as far away 
from the transition point in the  chaotic phase. These densities have not been studied in detail 
before. The newly introduced generalized Gauss map is particularly suitable for analytical 
calculations, as we will be able to solve the Perron-Frobenius operator fixed point equation 
at the critical point $\alpha_c$, as well as for large values of the parameter $\alpha$. 
We observe that our new map exhibits robust chaos for any $\alpha > \alpha_c$ as the Lyapunov 
exponent does not dip below zero  but stays positive (and, in fact, becomes arbitrarily large 
for $\alpha \to \infty$).


From a physical point of view, a generalized mixmaster universe dynamics described by our new 
map would abruptly change from a stable state into a chaotic one, staying chaotic 
for any $\alpha > \alpha_c$. From a renormalization group point of view, a stable fixed 
point of a renormalization group transformation described by this map would become abruptly chaotic. 
A generalized Brownian particle driven by the $\alpha$-Gauss map together with some friction 
force would switch abruptly from a constant deterministic velocity to a chaotic velocity at the 
critical point \cite{beck-physica-a}. We will provide arguments that some scaling features 
within this scenario, which we solve through analytic and numerical means, are universal, 
i.e., expected for other deterministic maps as well that have infinitely many symbols in their 
symbolic dynamics description.

\section{Fixed points of the $\alpha$-Gauss map}
\begin{figure}[h!]
\centering
\includegraphics[width=1.0\columnwidth]{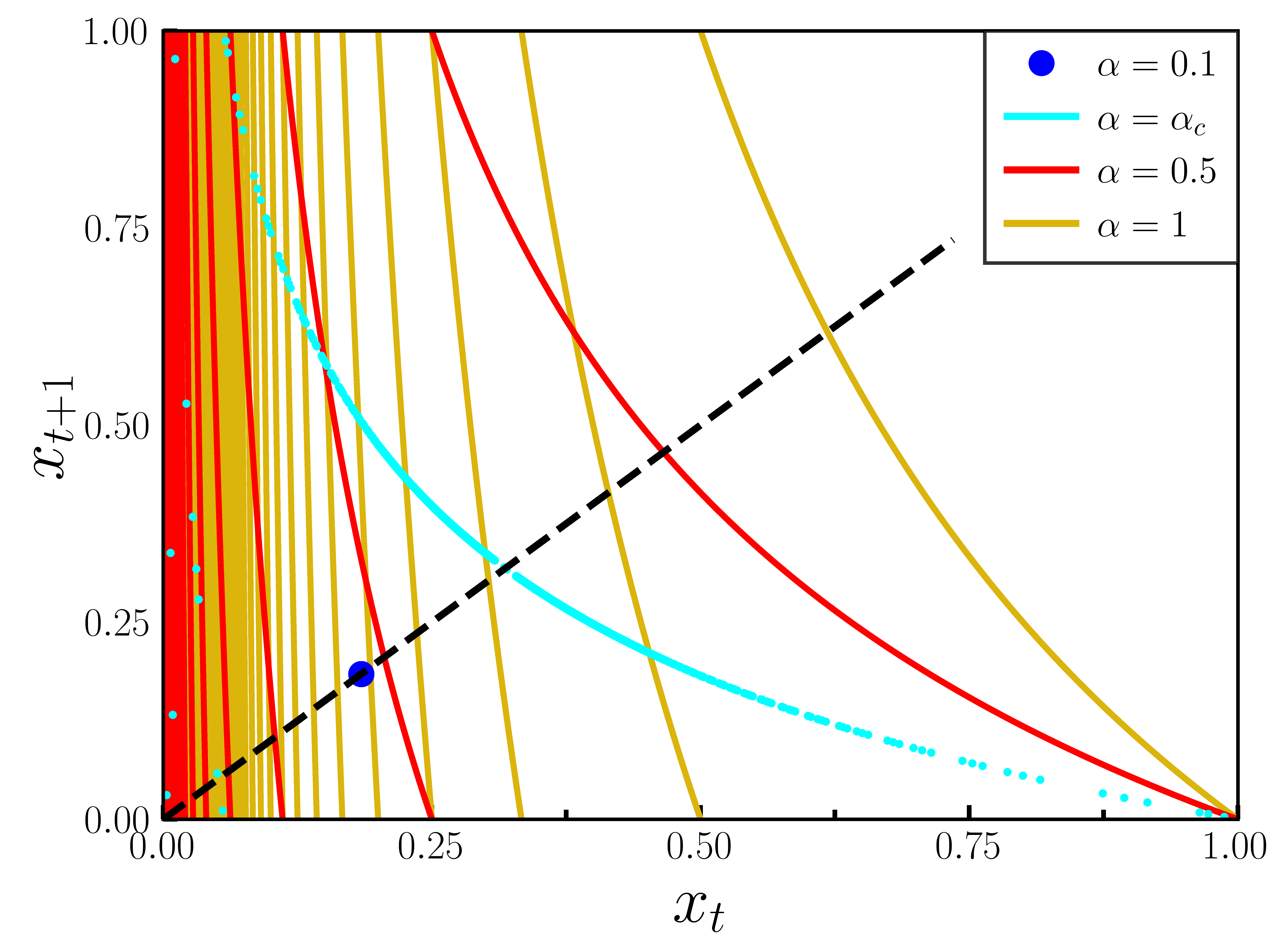}
\caption{Return map of the $\alpha$-Gauss map. For $\alpha<\alpha_c$ there is a stable fixed point, whereas for $\alpha>\alpha_c$ there is chaotic behaviour. The map consists of infinitely many branches.}
\label{fig:return}
\end{figure}

Inspired by the importance of long-range interactions in atomic, plasma and astrophysics, where the Coulomb and the gravitational interactions, and generally potentials of the form $V(x)\sim 1/x^\alpha$, play a crucial role, let us consider the one-dimensional map
\begin{equation} 
x_{t+1}=f(x_t)=\frac{1}{x_t^\alpha} - \Bigl[\frac{1}{x_t^\alpha} \Bigr] \,\,\,\,\,\,\,\,\;\;(\alpha \ge 0)
\label{alphaGauss}
\end{equation}
on the phase space $X$ given by the unit interval $X=[0,1]$ (we also define $f(0):=0$). 
To the best of our knowledge this map is new and has not been previously studied. Clearly, for $\alpha =1$,  it reduces to the well-known Gauss map, sometimes also called continued fraction map \cite{adler,mayer-roepstorff,BS}.
The graph of the map $f$ is shown for various values of $\alpha$ in Fig.~\ref{fig:return}.

As a first step of our analysis, let us determine all fixed points $x_*$ of this map.
The fixed point condition
\begin{equation}
    x_*=f(x_*)
\end{equation}
has infinitely many solutions, which we parametrize by the positive integer $n= [1/x_*^\alpha]$.
We may write
\begin{equation}
    x_*=\frac{1}{x_*^\alpha} -n
\end{equation}
or equivalently
\begin{equation}
    x_*=\frac{1}{(x_*+n)^\frac{1}{\alpha}}. \label{4}
\end{equation}    
where $n=1,2,3,4,5, \ldots$

\begin{figure}[h!]
\centering
\includegraphics[width=1.0\columnwidth]{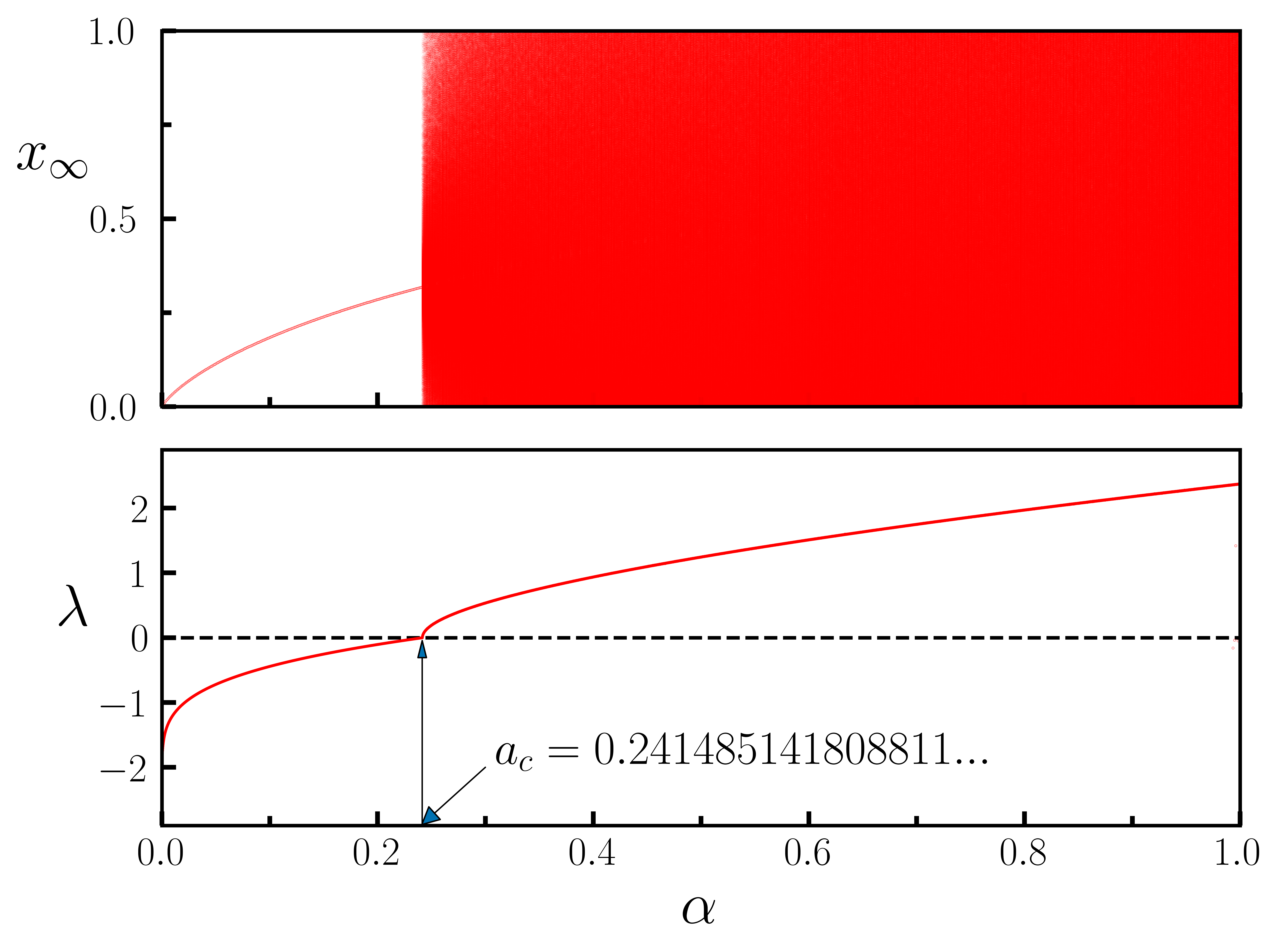}
\caption{Attractor ({\it top}) and Lyapunov exponent $\lambda$ ({\it bottom}) of the map. The sudden change from a stable fixed point to a  chaotic regime starting at the chaos threshold point $\alpha_c$ is evident. 
The behavior for $\alpha>1$ follows along the same lines; however, its numerical study requires a precision higher than the double precision used for $\alpha \in [0,1]$.}
\label{fig:bif_lya}
\end{figure}

\begin{figure}[h!]
\centering
\includegraphics[width=1.0\columnwidth]{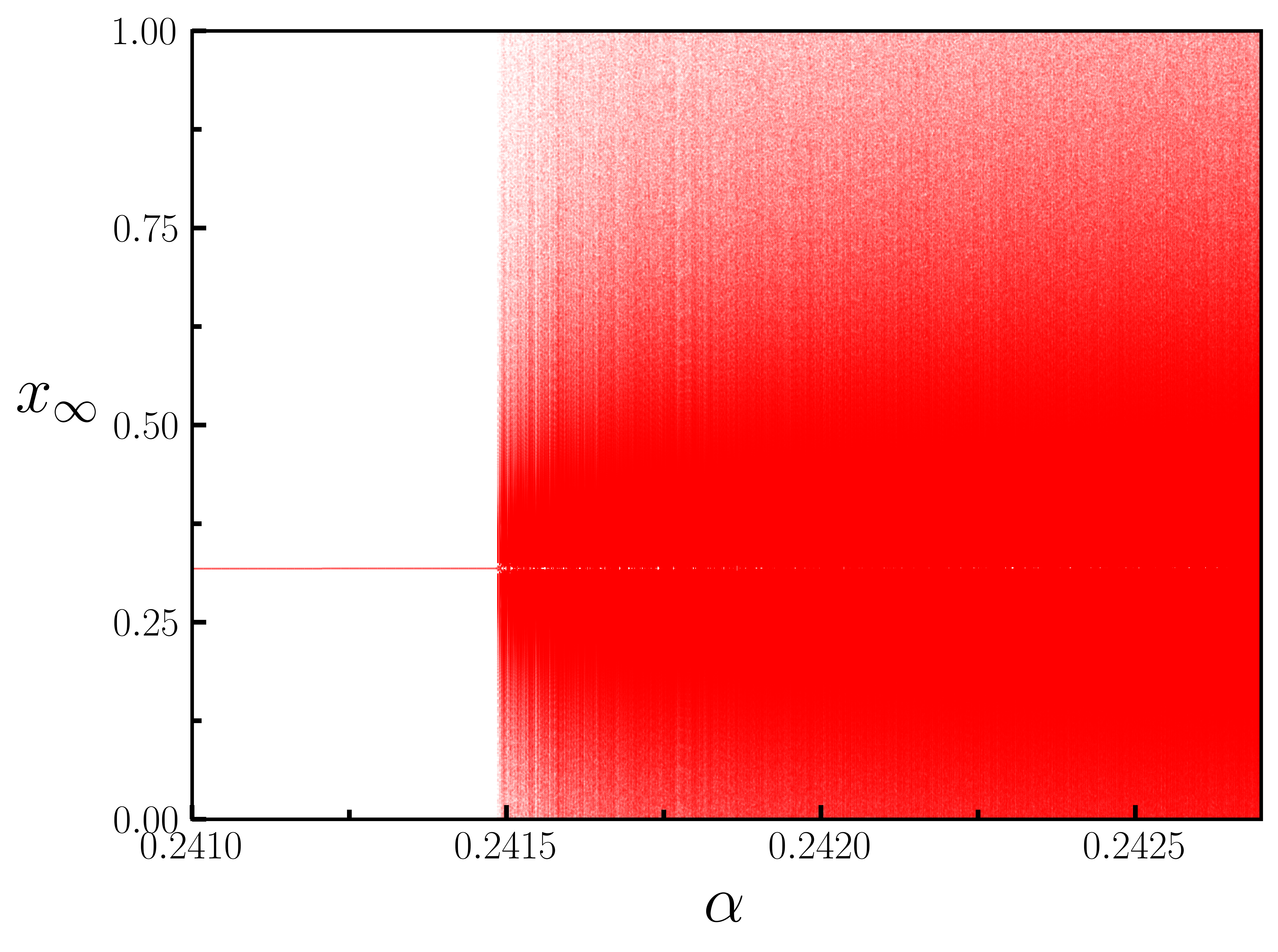}
\caption{Zoom of the attractor of the map in the vicinity of the chaos threshold $\alpha_c$.}
\label{fig:bif_lya_zoom}
\end{figure}

Next, we investigate the stability of the fixed points. Clearly the slope of the map $f(x)$ is negative for any value of $x$ such that
\begin{equation}
    f'(x)=-\alpha x^{-\alpha -1} \label{5}
\end{equation}
and, at a fixed point $x=x_*$, we may use Eq.~(\ref{4}) to write
\begin{equation}
    f'(x_*)=-\alpha (x_*+n)^\frac{\alpha+1}{\alpha}.
\end{equation}
Let us now look for critical parameter values $\alpha_c$, which
are parameter values where the stability of a given fixed point changes.
These are given by the condition
\begin{equation}
    f'(x_*)= -1.
\end{equation}
From Eq.~(\ref{5}) we obtain the statement that critical parameter values $\alpha_c$ must always satisfy
\begin{equation} 
-\alpha_c x_*^{-\alpha_c -1}=-1 \Longleftrightarrow x_*=\alpha_c^{\frac{1}{\alpha_c+1}} \label{8}
\end{equation}
We can now eliminate the fixed point $x_*$ in Eq.~(\ref{8}) by writing
\begin{equation}
    x_*=\frac{1}{x_*^{\alpha_c}}-n =\alpha_c^{-\frac{\alpha_c}{\alpha_c+1}}-n   \label{9}
\end{equation}
where $n= [x_*^{-\alpha_c}]=[\alpha_c^{-\frac{\alpha_c}{\alpha_c+1}}]$.
Combining Eqs.~(\ref{8}) and (\ref{9}) we get an implicit equation
for $\alpha_c$:
\begin{equation}
    \alpha_c^{\frac{1}{\alpha_c+1}}- \left( \frac{1}{\alpha_c}\right)^\frac{\alpha_c}{\alpha_c+1}+n=0 \label{10}
\end{equation}
For this equation (with $n=1,2,3,4,\ldots$), we have found numerically only one solution, namely that which corresponds to $n=1$:
\begin{equation}
\alpha_c=0.241485142 \dots
\end{equation}
The fixed point $x_*$ at this particular parameter value is given by
\begin{equation}
    x_*=\alpha_c^{\frac{1}{\alpha_c+1}}=0.318365736 \dots
\end{equation}
Figs.~\ref{fig:bif_lya} and \ref{fig:bif_lya_zoom} show the attractor of the $\alpha$-Gauss map as a function of $\alpha$. We see that the stable fixed point of the $\alpha$-Gauss map corresponding to $n=1$ looses its stability precisely at the above predicted point. All other fixed points (parametrized by $n\geq 2$) of the $\alpha$-Gauss map appear to be unstable for any $\alpha$, as Eq.~(\ref{10}) has no further solutions besides the one we found for $n=1$.

\begin{figure}[h!]
\centering
\includegraphics[width=.88\columnwidth]{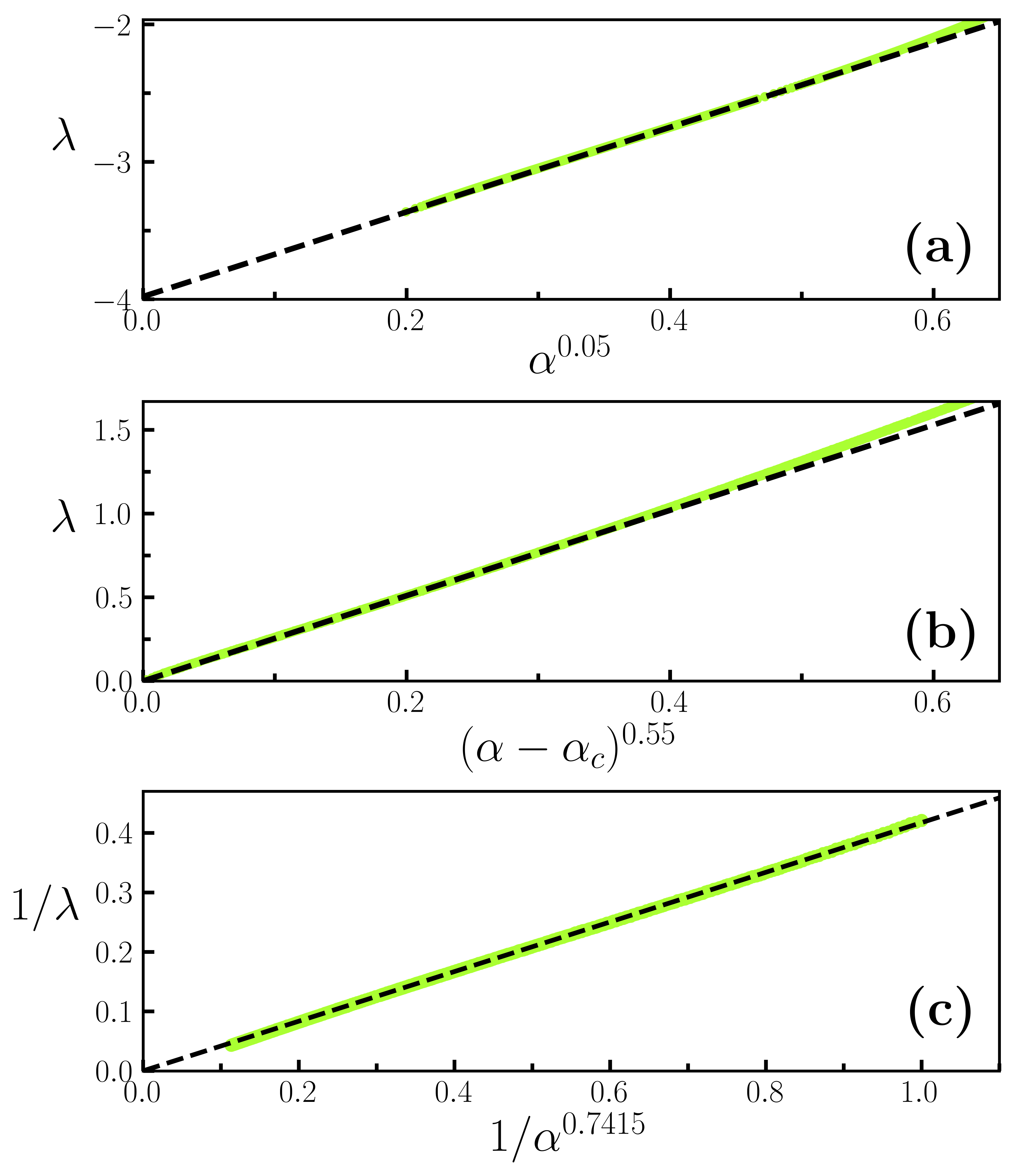}
\caption{(a)~Scaling behaviour of the Lyapunov exponent $\lambda$ near $\alpha =0$, indicating (dashed line) the possible asymptotic behavior $\lambda \sim -3.98 + 3.08\, \alpha^{0.05}\;(\alpha \to 0)$; 
(b)~Scaling of the Lyapunov exponent $\lambda$ near $\alpha_c$, indicating (dashed line) the asymptotic behavior $\lambda \sim 2.55 \left(\alpha-\alpha_c\right)^{0.55}\;(\alpha \to \alpha_c+0)$; {(c)~For $\alpha$ large enough, we numerically obtain $\lambda^{-1} \sim a /\alpha^b$ with $(a,b) \simeq (0.42,0.742)$}.}
\label{fig:lya_scale}
\end{figure}

In Fig. \ref{fig:lya_scale} we display the behaviour of the Lyapunov exponent in the neighborhood of $\alpha_c + 0$ and near $\alpha=0$. In the neighborhood of $\alpha=\alpha_c + 0$, we numerically obtain that $\lambda(\alpha) \propto (\alpha-\alpha_c)^{0.55}$. Moreover,
in the neighborhood of $\alpha=0$, we numerically obtain $\lambda(\alpha)-\lambda(0) \propto \alpha^{0.05}$. However, the exponent 0.05 being very close to zero, it cannot be numerically excluded that there is logarithmic rather than a power-law behaviour for $\alpha \to 0$. For large $\alpha$, there is numerical evidence that $1/\lambda \to 0$ for $\alpha \to \infty$.

\section{The invariant density of the $\alpha$-Gauss map}
\begin{figure}[h!]
\centering
\includegraphics[width=1.0\columnwidth]{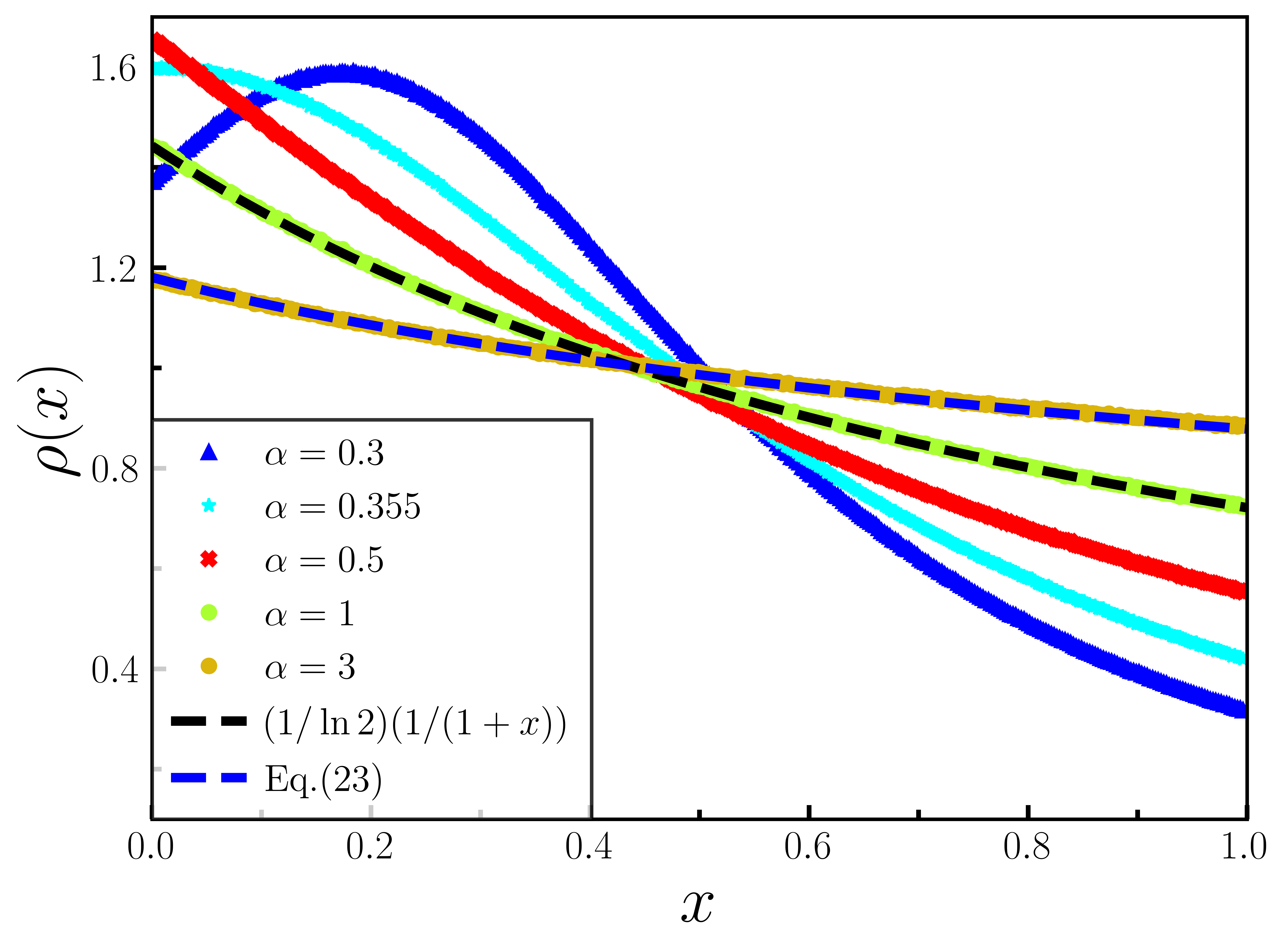}
\caption{Invariant density of the map for five representative values of the parameter $\alpha$. Note that $\alpha=1$ case corresponds to the standard Gauss map. It is evident from this plot that, as the chaos threshold $\alpha_c\approx 0.24$ is approached from above, the invariant density starts to develop a local maximum.}
\label{fig:Pvsx_a}
\end{figure}

Next, we wish to understand the invariant density of the $\alpha$-Gauss map in the chaotic regime. For this we have to solve the Perron-Frobenius fixed point equation for the density $\rho(y)$, given for general one-dimensional maps $f$ by
\begin{equation}
    \rho (y) = \sum_{x \in f^{-1}(y)}\frac{1}{|f'(x)|} \rho (x) \, .
\end{equation}
For the $\alpha$-Gauss map we have infinitely many pre-images $x$ for a given $y$, which are labelled by the integer $n$. As $f'(x)=-\alpha x^{-\alpha -1}$ we have
\begin{equation}
\frac{1}{|f'(x)|}=\frac{1}{\alpha} x^{\alpha +1}
\end{equation}
and the pre-images $x$ for a given $y$ are given by
\begin{equation}
    x= \frac{1}{(y+n)^\frac{1}{\alpha}} \;\;\;\; ( n=1,2,3, \ldots ) \, .
\end{equation}
Consequently, the invariant density $\rho(y)$, for arbitrary $\alpha$, satisfies the functional equation
\begin{eqnarray}
    \rho (y) &=& \frac{1}{\alpha} \sum_{n=1}^\infty x_n^{\alpha +1} \rho (x_n) \nonumber \\
    &=& \frac{1}{\alpha} \sum_{n=1}^\infty \frac{1}{(y+n)^\frac{\alpha+1}{\alpha}}\rho \left( \frac{1}{(y+n)^\frac{1}{\alpha}} \right) \, . \label{16}
\end{eqnarray}
Fig.~\ref{fig:Pvsx_a} shows some numerically obtained invariant densities (from long-term iteration of the map) for the example parameters $\alpha=0.3, 0.355, 0.5$, $1$ and $3$. By construction these densities must satisfy the functional equation (\ref{16}).

\begin{figure}[h!]
\centering
\includegraphics[width=0.7\columnwidth]{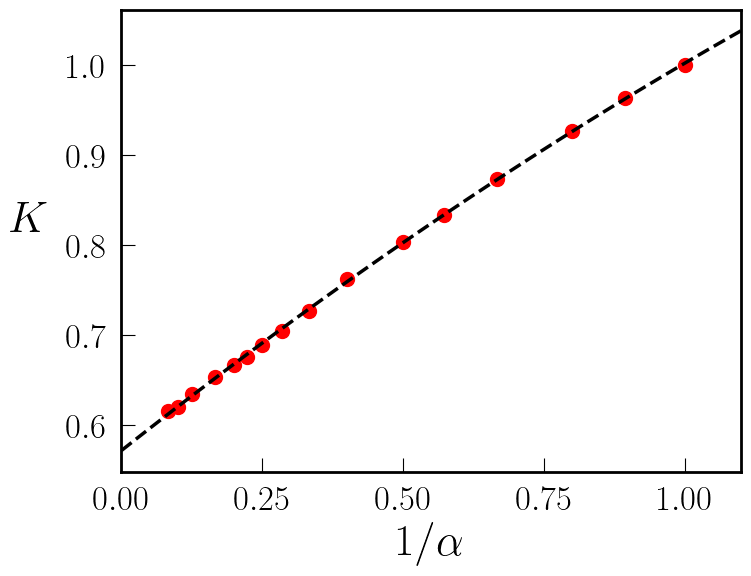}
\caption{{$\alpha$-dependence, for $\alpha$ large enough, of the parameter $K$ in Eq.~(\ref{heuristic}). We numerically verify $K \sim c+ d/\alpha$ with $(c,d) \simeq (0.57,0.50)$.}}
\label{Kalpha}
\end{figure}


Let us here consider a few special cases of $\alpha$ where we can solve the Perron-Frobenius fixed point equation explicitly.

The first case is $\alpha =1$, corresponding to the ordinary Gauss map. Here it is known \cite{adler, mayer-roepstorff, BS} that the invariant density is given by
\begin{equation}
    \rho(y)= \frac{1}{\ln 2} \,\frac{1}{1+y} \,\, .\label{17}
\end{equation}
One can check that this density indeed satisfies Eq.~(\ref{16}) for $\alpha =1$. To this end, we write the left-hand side of Eq.~(\ref{16}) as
\begin{equation}
    l= \frac{1}{\ln 2} \, \frac{1}{1+y} \label{18}
\end{equation}
and its right-hand side as
\begin{eqnarray}
    r &=& \frac{1}{\ln 2} \sum_{n=1}^\infty \frac{1}{(y+n)^2} \, \frac{1}{1+\frac{1}{y+n}} \nonumber \\
    &=& \frac{1}{\ln 2} \sum_{n=1}^\infty \frac{1}{(y+n)(y+n+1)} \, \,\, .\label{19}
\end{eqnarray}
One can then easily check (for example, by using a symbolic computation program) that $l=r$, i.e., the infinite sum in Eq.~(\ref{19}) adds up to what is written in Eq.~(\ref{18}).

The second case where we can solve the Perron-Frobenius equation exactly is the critical case $\alpha = \alpha_c=0.241485141...\,$ Here the solution of Eq.~(\ref{16}) is simply
the Dirac delta function
\begin{equation}
    \rho(y) =\delta (y-y_*) \label{20}
\end{equation}
where $y_*=\alpha_c^\frac{1}{\alpha_c+1}=0.318365736...$ is the fixed point of the map that looses stability precisely at this critical point.

To see this, we note that the Dirac delta function $\delta(y-y_*)$ vanishes everywhere except for $y=y_*$ which is the fixed point that we labelled with $n=1$.
Hence, Eq.~(\ref{16}) reduces to a sum over just one value, namely $n=1$. We have
\begin{equation}
    \rho (y) = \frac{1}{\alpha}\, \frac{1}{(y+1)^\frac{\alpha+1}{\alpha}} \,\,
    \rho \left( \frac{1}{(y+1)^\frac{1}{\alpha}} \right) .
\end{equation}
But, since $y=y_*=1/(y_*+1)^\frac{1}{\alpha}$ is the fixed point, the density $\rho$ on the left-hand side and right-hand side evaluated at the fixed point $y=y_*$ is the same, meaning we end up with the condition
\begin{equation}
    \frac{1}{\alpha}\, \frac{1}{(y_*+1)^\frac{\alpha +1}{\alpha}}=1 \,\, .
\end{equation}
This is just the formula that we already previously derived for the $n=1$ fixed point in section II, see Eq.~(\ref{9}) in that section. Therefore, the Dirac delta as given by Eq.~(\ref{20}) is indeed a solution to the Perron-Frobenius equation at precisely the critical parameter $\alpha=\alpha_c$. 

We notice here a deeper mechanism underlying our `jump to chaos' scenario. A fixed point $y_*$ of the map itself becomes a fixed point of the corresponding Perron-Frobenius operator via the function $\delta (y-y_*)$; this happens precisely at the critical value $\alpha_c$. Then, if $\alpha$ moves slightly away from $\alpha_c$, the Dirac delta peak smoothens if $\alpha>\alpha_c$. Chaotic behavior sets in immediately.

Finally, it is also possible to analytically calculate the invariant density, in good approximation, for large values of $\alpha$. The final result of a calculation described in detail in section VI is that $\rho(y)$ is approximately given by

\begin{equation}
 \rho(y) = \frac{C}{ (K+y)^{1/\alpha}}, 
\label{heuristic} 
\end{equation}
where $C=C(\alpha)=b/((K+1)^b - K^b)$ is the normalization constant, $b=b(\alpha)=1-1/\alpha$, and $K=K(\alpha)$ is a fitting parameter that depends smoothly on $\alpha$, as displayed in Fig.~\ref{Kalpha}.
We verify that, in the limit $\alpha \to 1$, we recover from Eq.~(\ref{heuristic}) the exact density Eq.~(\ref{17}).
 It turns out that this approximation is a good one for any $\alpha\ge 1$ (see Fig.~5). In the limit $\alpha\to\infty$, we obtain $\rho(y)=1$, which is numerically verified.

\section{Behavior for $\alpha$ slightly above $\alpha_c$}
\begin{figure}[h!]
\centering
\includegraphics[width=1.0\columnwidth]{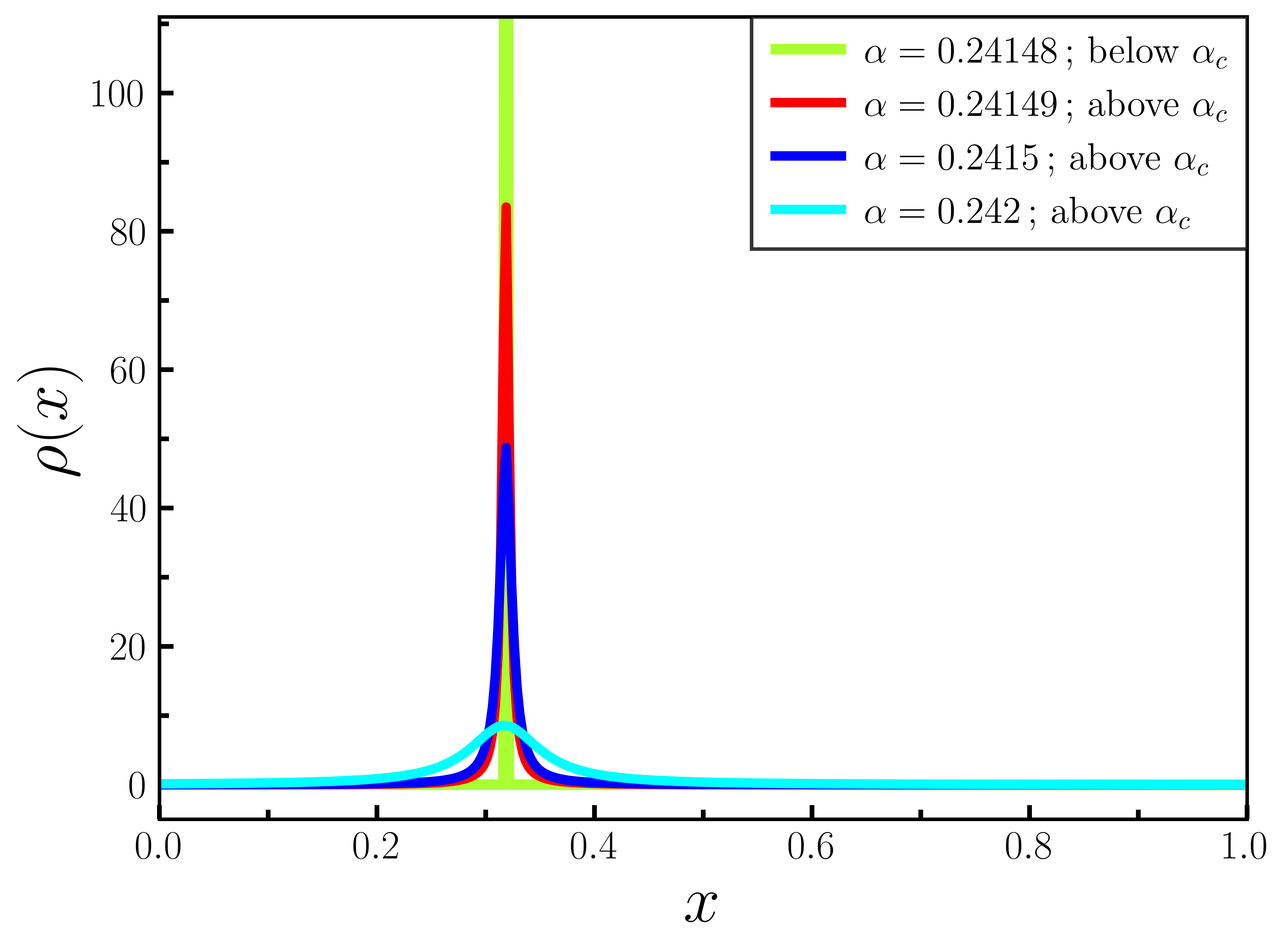}
\caption{Invariant density of the map for three values of the parameter $\alpha$ gradually approaching the point $\alpha_c$ from above as well as one value of $\alpha$ which is slightly below $\alpha_c$. 
}
\label{fig:Pvsx}
\end{figure}

\begin{figure}[h!]
\centering
\includegraphics[width=1.0\columnwidth]{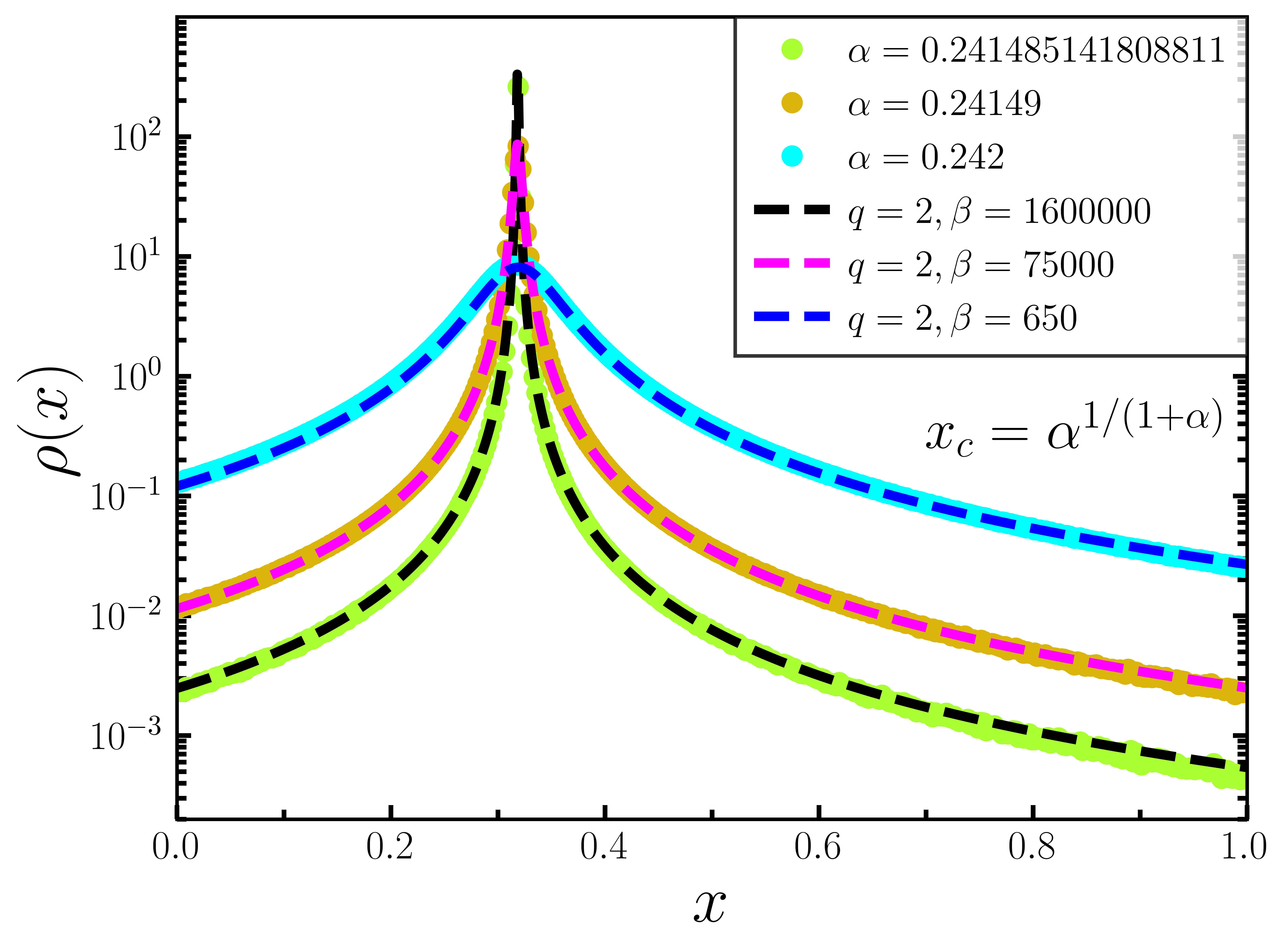}
\caption{Invariant density of the map for 3 values of the parameter $\alpha$
slightly above $\alpha_c$. It is evident that each curve is well approximated by a $q$-Gaussian with $q = 2$, namely $P(x)=\frac{\sqrt{\beta}}{\pi}\frac{1}{1+\beta (x-x_c)^2}$ .}
\label{fig:Pvsx_fit}
\end{figure}

Numerically, we observe the interesting phenomenon that, if the parameter $\alpha$ is very close but slightly above the critical value $\alpha_c$, then the invariant density is well-fitted by a $q$-Gaussian with $q=2$ 
(i.e. a Cauchy distribution) that is very sharply peaked around the fixed point $y_*$. This is illustrated in Figs.~\ref{fig:Pvsx}   and \ref{fig:Pvsx_fit}. The fitting formula that we used is
\begin{equation}
    \rho(y) =\frac{\sqrt{\beta}}{\pi} \frac{1}{1+\beta (y-y_*)^2}
\end{equation}
where $\beta$ is very large so that the density is very sharply peaked
and approximating the Dirac delta for $\beta \to \infty$. 

\begin{figure}[h!]
\centering
\includegraphics[width=1.0\columnwidth]{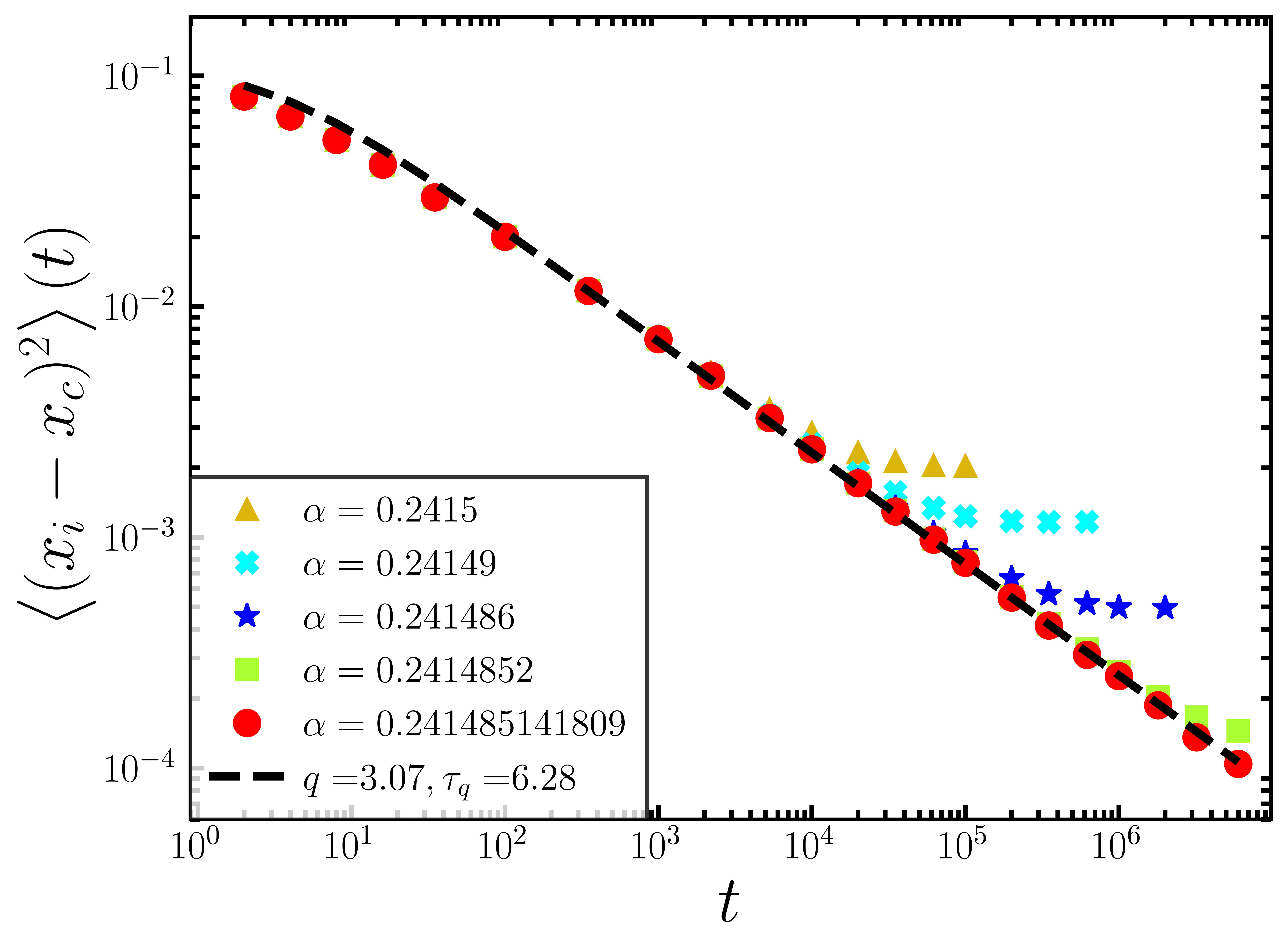}
\caption{ Decreasing width of the distribution as a function of time $t$ if the uniform density is chosen as initial distribution.
$N$ is the number of boxes of the histogram chosen for the calculation (we used $N=400$). It is evident that a $q$-exponential with a power-law tail in time develops 
as $\alpha$ gets closer to the exact critical value $\alpha_c$. For the closest $\alpha$ value that we used, up to very large time steps ($\sim 10^7$) no deviation from the power law is observed.  
 }
\label{fig:relax}
\end{figure}

If, at $t=0$, we start with the uniform density, the width of the time-dependent density in this numerical experiment gradually decreases with time as depicted in Fig.~\ref{fig:relax}. As $\alpha$ approaches $\alpha_c$ from above, the width around the value $x_c$ is numerically observed to approach zero as follows:
\begin{equation}
\langle (x -x_c)^2 \rangle(t)= \langle (x -x_c)^2 \rangle(0)\, e_q^{-t/ \tau_q} = \frac{ \langle (x -x_c)^2 \rangle(0)}{[1+(q-1)t/\tau_q]^{1/(q-1)}}
\label{tauq}
\end{equation}
with $(q,\tau_q) \simeq (3.1, 6.3)$ and
\begin{equation}
\langle (x -x_c)^2 \rangle(0)= \int_0^1 dx (x-x_c)^2 = \frac{1}{3}-x_c + x_c^2=0.116 \dots 
\end{equation}
This behavior, as $t\to\infty$, is strongly reminiscent of the $q$-generalized Large Deviation Theory that is observed, as the number of random variables $N\to\infty$, in various models with $q$-Gaussian invariant densities \cite{RuizTsallis2012,Touchette2013,RuizTsallis2013,TirnakliTsallisAy2021,TirnakliMarquesTsallis2022,ZamoraTsallis2022}.

\begin{figure}[h!]
\centering
\includegraphics[width=.88\columnwidth]{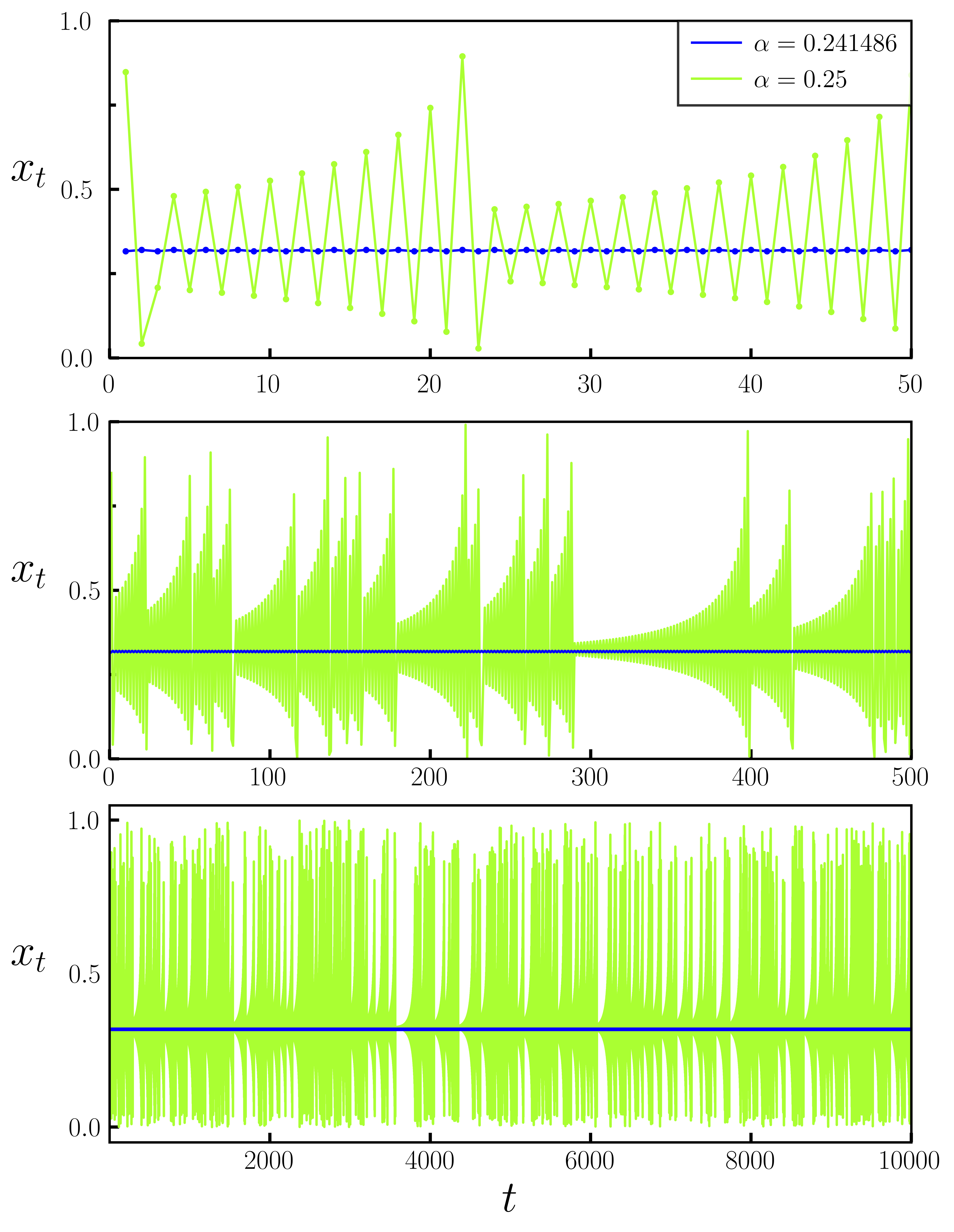}
\caption{Chaotic trajectories as a function of time for the parameter value $\alpha=0.25$ which is slightly above $\alpha_c$. After random injection close to the unstable fixed point, the distance to the unstable fixed point increases in an exponential way for a limited amount of time, until there is re-injection again.}
\label{fig:xvst}
\end{figure}

The chaotic dynamics for $\alpha$ slightly above $\alpha_c$ is of a very special type, 
as illustrated in Fig.~\ref{fig:xvst}. Trajectories move slowly away from the unstable 
fixed point $y_*$ and are quasi-randomly re-injected towards the vicinity of the fixed point. 
This repeats in an intermittent way (see \cite{6} for a summary of possible 
intermittency scenarios for different classes of maps). 
The Lyapunov exponent is positive but very close to zero (see numerical results in 
Figs.~\ref{fig:bif_lya} and \ref{fig:lya_scale}).

\section{Universality of the density at the critical point}

Let us finally provide some empirical arguments 
to derive that the value of the entropic index is $q=2$ for the
invariant density close to the critical point in our transition scenario,
meaning we have a Cauchy distribution for $\alpha$ near $\alpha_c+0$. At the same time we
will provide some arguments for the universality of the observed
scenario.

Suppose we are slightly above the critical value $\alpha_c$ where the stable
fixed point of the $\alpha$-Gauss map, or of a similar map with infinitely many branches, has lost its stability but still persists as an unstable
fixed point. The invariant density is denoted by $\rho(x)$ and if we
numerically determine a histogram of iterates with bin size $\Delta$ we have
\begin{equation}
    p_i = \int_{i\Delta}^{(i+1)\Delta} \rho(x')dx'
\end{equation}
for the probability $p_i$ to find the iterate in the $i$-th bin. For $\alpha$ slightly above $\alpha_c$
the trajectory exhibits the special dynamics displayed in Fig.~\ref{fig:xvst}. After a random
injection close to the unstable fixed point the distance of the trajectory to the unstable fixed point increases exponentially for a while, until it is random-injected again. Let us now determine the time $t$ that the trajectory spends in a given bin of size $\Delta$ close to the fixed point. If injected at distance $x$ from the fixed point and if we assume that $x$ coincides with
the left boarder of a given bin interval then $t$ is determined by the condition
\begin{equation}
    x \, e^{\lambda t}=x+\Delta,
\end{equation}
since after time $t$ the right border of the bin interval is reached. Here $\lambda$ is the (very small positive) Lyapunov exponent of the map.
From the above we get 
\begin{equation}
    \lambda \, t= \ln \left(1+\frac{\Delta}{x}\right) \approx \frac{\Delta}{x}
\end{equation}
if $x >> \Delta$. Now the time spent in the bin of size $\Delta$
is proportional to the probability $p_i$ since we can determine the invariant density as a time average.
Thus
\begin{equation}
    p_i \sim t \sim \frac{\Delta}{\lambda x} \sim \frac{1}{x}
\end{equation}
This means the probability $p_i$ is inversely proportional to $x$,
the distance from the unstable fixed point. Thus the probability
density $\rho$ must be proportional to $1/x^2$. Since $x=y-y_*$ is the distance from the unstable fixed point $y_*$ we thus get
\begin{equation}
    \rho (y) \sim (y-y_*)^{-2} 
\end{equation}
which implies $q=2$, since a $q$-Gaussian decays with the power-law exponent
$-\frac{2}{q-1}$.

We notice that our argument only assumed the existence of the unstable fixed point $y_*$ and a random injection close to the fixed point, created by the infinitely many branches of the map close to $y=0$. Our argument does not
depend on any other details of the map considered. Hence we have reason to believe that the derived $q=2$-Gaussian shape of the invariant density close to the critical point is universal, meaning it occurs quite generically for the type of transition scenarios considered here.

\section{Behaviour for large $\alpha>1$}

For $\alpha >1$ we observe that the chaotic attractor persists for any value of $\alpha$. There are no stable periodic windows in this region (this is very different from, e.g., the scenario for the logistic map). We numerically observe chaos and only chaos for any $\alpha > \alpha_c$. This distinguishes our scenario from previously studied scenarios of other maps. The Lyapunov exponent $\lambda$ is positive for any $\alpha>\alpha_c$ and it approaches infinity if $\alpha \to \infty$ (see numerical results for $1/\lambda$ in Fig.~\ref{fig:lya_scale}c, which provide numerical indication of a scaling law as $1/\lambda \to 0$). It is remarkable how smoothly $\lambda$ depends on $\alpha$, which is again very different from the logistic map. 

We should remark here that the numerical investigation of the $\alpha$-Gauss map for $\alpha >1$ is very delicate. The quantity $1/x^\alpha$ can become huge and one has to take extra care that the integer value $[1/x^\alpha]$ is correctly evaluated (it can be a huge number) such that the difference $1/x^\alpha-[1/x^\alpha]$ is correctly calculated. In our numerical evaluation of the map we have taken that extra care.
While the case of large $\alpha$ is apparently numerically much more difficult for the correct numerical iteration of the map, it is remarkable that analytically it leads to a simplification for the calculation of the invariant density. This we will show in the following.

Assume some arbitrary initial density $\rho_j(y)$ on the interval [0,1] is given, then in the next iteration step of the map this density changes to $\rho_{j+1}(y)$ given by
\begin{eqnarray}
    \rho_{j+1} (y) &=& \frac{1}{\alpha} \sum_{n=1}^\infty x_n^{\alpha +1} \rho_j (x_n) \nonumber \\
    &=& \frac{1}{\alpha} \sum_{n=1}^\infty \frac{1}{(y+n)^\frac{\alpha+1}{\alpha}}\rho_j \left( \frac{1}{(y+n)^\frac{1}{\alpha}} \right) \, . \label{66}
\end{eqnarray}
This is just the 1-step iteration of the Perron-Frobenius operator for the $\alpha$-Gauss map, see section III. The exact invariant density is obtained if the above equation is iterated infinitely often, i.e.\ in the limit $ j \to \infty$ we obtain the invariant density for an arbitrary initial distribution $\rho_0$ that is smooth \cite{BS}. However, 
for exponentially mixing maps a good approximation is often already obtained after very few iterations.
Let us choose in zeroth-order approximation for large $\alpha$ the initial density $\rho_0(y)=1$. Then a better approximation of the invariant density is given if we apply the Perron-Frobenius operator once to this initial density $\rho_0(y)=1$, obtaining
\begin{equation}
    \rho_{1} (y) 
    = \frac{1}{\alpha} \sum_{n=1}^\infty \frac{1}{(y+n)^\frac{\alpha+1}{\alpha}} \, . \label{67}
\end{equation}
While in principle one should continue the iteration process,
a 1-step iteration may already be a good approximation if we start from an initial density such as $\rho_0$ that is already close to the invariant density.
We can now approximate the infinite sum in Eq.~(\ref{67}) by an integral and obtain therefore
an approximate expression for the invariant density $\rho(y) \approx \rho_1(y)$. As the sum  starts at $n=1$, we allow the integral to start at some value $K$ which should be close to 1 but we allow it to be fitted in the best possible way for each $\alpha$, so that the sum is replaced by an integral starting at $K$. From this we get the following approximate formula for the
invariant density
\begin{equation}
\rho (y) \approx \frac{1}{\alpha} \int_K^\infty \frac{1}{(y+x)^{1+\frac{1}{\alpha}}} dx =:I
\end{equation}
The integral $I$ can be analytically calculated and we get
\begin{equation}
I= (K+y)^{-\frac{1}{\alpha}}.    \label{35}
\end{equation}
Remarkably, if we choose the special case $\alpha =1$ and choose $K=1$,
we get the functional dependence $1/(1+y)$ which is just the $y$-dependence of the exact invariant density of the usual Gauss map. Thus, although we did several approximations, these compensate and in the end we get the exact expression for the Gauss map for $\alpha =1$. Thus the expression derived is also expected to be a good approximation for other values of $\alpha$, in particular $\alpha >1$ where the relaxation for equilibrium becomes very fast. We still need to normalize the distribution such 
that $\rho(y)$ satisfies $\int_0^1 \rho(y)dy=1$. From this we arrive at the final result
\begin{equation}
    \rho(y)\approx \frac{C}{(K+y)^{\frac{1}{\alpha}}}
\end{equation}
where
\begin{equation}
    C=\frac{b}{(K+1)^b-K^b} \,\,\,\,\,\, \mbox{with } \,\,\,\, b:=1-\frac{1}{\alpha}.
    \end{equation}
This analytically derived result agrees quite well with the numerically obtained histogram of 
the iterates of the map for any $\alpha >1$, see Fig.~{\ref{fig:Pvsx_a}}. 
For $\alpha \to \infty$ we get $\rho(y)=1$.

\section{Concluding remarks}
In this paper we have introduced and analysed a generalization of the traditional Gauss map. 
The definition (\ref{alphaGauss}) generates a new  family of non-invertible 1-dimensional maps 
on the interval which depend on a parameter $\alpha>0$ and which have infinitely many pre-image 
branches for any given $\alpha>0$. The usual Gauss map is recovered for $\alpha =1$. 
A combination of analytical and numerical results reveals an intriguing scenario, a road to 
chaos which we may call a {\it `jump into chaos'} (see \cite{1,2} for other types of maps that 
have been numerically observed to exhibit a similar transition). The interesting and novel 
properties of this scenario are as follows: The Lyapunov exponent $\lambda$ monotonically and 
continuously increases from negative values to +infinity when the control parameter $\alpha$ 
increases from zero to infinity. Thus the end point of our scenario is a state with infinite 
Lyapunov exponent--rather than just a finite positive value as is the case for the logistic map. 
Moreover, the scenario occurs in a rather simple and monotonic way---there are no bifurcations, 
no periodic windows within the chaotic regime, there is just a stable period-1 orbit for small 
$\alpha$ which abruptly switches to a chaotic attractor at the critical value 
$\alpha_c=0.241485142 \dots$. 
Above the critical threshold, chaos persists for any $\alpha$ (no interruption by stable periodic 
windows, an example of robust chaos \cite{8,9} but here exhibited by a new type of 
map with infinitely many branches).
The function $\lambda (\alpha)$ crosses zero and is exhibiting a singularity in its derivative 
at the critical point $\alpha_c$, otherwise it is a smooth and monotonic function over the whole 
range of $\alpha \in [0, \infty]$.

The invariant density of the system can be understood by analytical means. For $\alpha<\alpha_c$ 
there is a stable fixed point, hence the invariant density is a Dirac delta function centered 
around this fixed point. For $\alpha >\alpha_c$, there is chaotic behaviour and a smooth  
invariant density is observed, whose shape depends on $\alpha$. The invariant density of the 
ordinary Gauss map ($\alpha =1$) is exactly known, but we also derived approximate formulas 
for the invariant density for other values of $\alpha$, which are a good approximation for 
larger values of $\alpha$. For $\alpha \to \infty$, the uniform density on [0,1] is obtained. 
Precisely at the critical value $\alpha_c$, $q$-statistical features emerge. The invariant 
density, in good approximation, is a $q$-Gaussian with $q=2$ at the critical point (in other words, 
it is a Cauchy distribution) and the width of this distribution becomes infinitely narrow as the 
stable fixed point is approached. We also investigated how an initially uniform distribution 
evolves in time at the critical point $\alpha =\alpha_c$, and here we numerically observe that 
the time development of the approach to equilibrium is described by a $q$-exponential with 
$q \simeq 3.07$.

One may conjecture that some sort of universality exists for a wider class of deterministic maps 
that have infinitely many symbols in their symbolic dynamics description. The $\alpha$-Gauss map 
studied here was our main example, but one may construct other examples as well. 
Maps with infinitely many branches (or pre-images), as studied here, apparently follow 
different routes to chaos than in the ordinary period-doubling, intermittency or quasi-periodic 
scenarios. These kinds of maps allow for a transition route where the Lyapunov exponent becomes 
not just positive but can become infinitely large at the end point of the scenario---a situation 
that has not been studied much before. It is very interesting to see that for our maps studied 
here there is precisely one critical point $\alpha_c$, with chaos above $\alpha_c$, and a stable 
period 1 below $\alpha_c$, and nothing else, a true `jump into chaos' when $\alpha$ crosses the 
threshold. This situation is very different from e.g. the logistic map. 

Our derivation of the Cauchy invariant density ($q=2$) at the critical point only required 
the existence of a fixed point of the map which changes its stability, with trajectories 
quasi-randomly injected to the vicinity of the unstable fixed point of the map by the infinitely 
many branches of the map if the control parameter is slightly above the critical value. 
Hence we expect this $q=2$ result to be universal, i.e. it should occur in a similar way for 
other maps that have infinitely many branches and for which a single critical point exists 
where the Lyapunov exponent vanishes.

Another interesting feature of our scenario is the end point of the route to chaos where the 
Lyapunov exponent appears to diverge to +infinity, in the limit $\alpha \to \infty$. 
Here the dynamics becomes extremely chaotic,
and we may refer to this state as {\em extreme chaos}. There is extreme sensitivity on 
initial conditions, and the numerical investigation becomes very difficult as extremely 
high precision is needed for each iteration step. The deterministic map basically degenerates 
to a dense system of infinitely many branches where each one has infinite slope. But an analytic 
treatment is still possible. As we have shown, the invariant density of this deterministic 
system just becomes the uniform density on the unit interval. However, this state is even 
more random than the well-known
Bernoulli shift of the binary shift map, which just has Lyapunov exponent $\ln 2$ \cite{yan2020}. 
In our case, we have infinite positive Lyapunov exponent due to the infinitely many branches of 
infinite slope.




\begin{acknowledgments} 
The numerical calculations reported in this paper were partially performed at TUBITAK ULAKBIM, 
High Performance and Grid Computing Center (TRUBA resources). U.T. is a member of the Science 
Academy, Bilim Akademisi, Turkey and supported by the Izmir University of Economics Research 
Projects Fund under Grant No. BAP-2024-07. 
C.T. is partially supported by CNPq and Faperj (Brazilian agencies). C.B. is supported by an 
ISPF-ODA grant of QMUL. He is also supported by an International Excellence Fellowship of KIT 
Karlsruhe.
\end{acknowledgments}




\begin{thebibliography}{99}

\bibitem{adler} R.L. Adler and L. Flatto, Ergod. Th. Dyn. Systems {\bf 4}, 487 (1984).

\bibitem{sinai} Y.G. Sinai and C. Ulcigrai, Ergod. Th. Dyn. Systems {\bf 28}, 643 (2008).

\bibitem{corless} R.M. Coreless, Am. Math. Mon. {\bf 99}, 203 (1992).

\bibitem{banchoff}
T. Banchoff, T. Gaffney, C. McCrory and D. Dreibelbis, {\it Cusps of Gauss Mappings}, Research Notes in Mathematics {\bf 55} (Pitman Publisher, London, 1982) 

\bibitem{BS} C. Beck and F. Schloegl, {\em Themodynamics of Chaotic Systems}, Cambridge University Press 
(1993).

\bibitem{mayer-roepstorff} D. Mayer and G. Roepstorff, J. Stat. Phys. {\bf 47}, 149 (1987).

\bibitem{chernoff} D.F. Chernoff and J.D. Barrow, Phys. Rev. Lett. {\bf 50}, 134 (1983).

\bibitem{barrow} J.D. Barrow, Phys. Rev. D {\bf 102}, 024017 (2020).

\bibitem{lacasa} J. Calero-Sanz, B. Luque, L. Lacasa, Phys. Rev. E {\bf 107}, 044214 (2023).


\bibitem{grebogi} C. Grebogi, E. Ott, J.A. Yorke, Physica D {\bf 7}, 181 (1983).

\bibitem{ott} E. Ott, Scholarpedia {\b 1}(10): 1700 (2006).

\bibitem{thomson} J.M.T. Thomson, H.B. Stewart, Y. Ueda, Phys. Rev. E {\bf 49}, 1019 (1994).

\bibitem{luo} K.J. Luo, H.T. Grahn, K.H. Ploog, L.L. Bonilla, Phys. Rev. Lett. {\bf 81}, 1290 (1998).

\bibitem{yue} X. Yue, W. Xu, L. Wang, Commun. Nonl. Sci. Num. Sim. {\bf 18}, 3567 (2013).



\bibitem{1} T. Kawabe, K. Kondo, Prog. Theor. Phys. {\bf 85}, 759 (1991).

\bibitem{2} O. Alvarez-Llamoza, M. G. Cosenza, G. A. Ponce, Chaos Solitons 
Fractals {\bf 36}, 150 (2008).

\bibitem{3} T. Kawabe, K. Kondo, Prog. Theor. Phys. Progress Letters {\bf 86}, 581
(1991).

\bibitem{4} A.S. Pikovsky, J. Phys. A: Math. Gen. {\bf 16}, L109-LI12 (1983).

\bibitem{5} M.G. Cosenza, O. Alvarez-Llamoza, G.A. Ponce, Comm. Non. Sci.
Num. Sim. 1{\bf 5}, 2431 (2010).

\bibitem{6} H.G. Schuster, W.Just, {\em Deterministic Chaos, An Introduction},
Wiley‐VCH Verlag (2005).


\bibitem{7} J.M. Aguirregabiria, arXiv:0907.3790 (2009).

\bibitem{8} S. Banerjee, J.A. Yorke, C. Grebogi, Phys. Rev. Lett. {\bf 80}, 3049
(1998).

\bibitem{9} E. Zeraoulia, J.C. Sprott, {\em Robust Chaos and Its Applications},
World Scientific (2011).

\bibitem{10} M. Andrecut, M.K. Ali, Phys. Rev. E {\bf 64}, 25203 (2001).

\bibitem{11} J.A.C. Gallas, Int. J. Bifur. Chaos {\bf 20}, 197 (2010).

\bibitem{beck-physica-a} C. Beck, Physica A {\bf 233}, 419 (1996).

\bibitem{TirnakliBorges} U. Tirnakli and E.P. Borges, Sci. Rep. {\bf 6}, 23644 (2016).


\bibitem{RuizTsallis2012}G. Ruiz and C. Tsallis, 
Phys. Lett. A  {\bf 376}, 2451-2454 (2012).

\bibitem{Touchette2013}H. Touchette, 
Phys. Lett. A {\bf 377} (5), 436-438 (2013).

\bibitem{RuizTsallis2013}G. Ruiz and C. Tsallis, 
Phys. Lett. A {\bf 377}, 491-495 (2013).

\bibitem{TirnakliTsallisAy2021}U. Tirnakli, C. Tsallis and N. Ay, 
Nonlinear Dynamics {\bf 106}, 2537-2546 (2021).

\bibitem{TirnakliMarquesTsallis2022}U. Tirnakli, M. Marques and C. Tsallis, 
Physica D {\bf 431}, 133132 (2022).

\bibitem{ZamoraTsallis2022}D.J. Zamora and C. Tsallis, 
Physica A  {\bf 608},  128275 (2022).

\bibitem{yan2020} J. Yan and C. Beck, Chaos Solitons Fractals X {\bf 005}, 100035 (2020).

\end{thebibliography}
\end{document}